\documentstyle[aps,multicol]{revtex} 
\def\be{\begin{equation}} 
\def\ee{\end{equation}} 
\def\ho{{\hbar\omega}} 
\def\shat{{\hat\sigma}} 
\def\ahat{{\hat a}} 
\def\pr{{\prime}} 
\begin{document} 
\draft 
\title{Quantum Computation with Quantum Dots and Terahertz Cavity 
Quantum Electrodynamics} 
\author{Mark S. Sherwin$^{1,2}$, Atac Imamoglu$^{2,3}$ and Thomas 
Montroy$^{1}$} 
\address{ 
$^{1}$ Physics Department, University of California 
Santa Barbara, CA 93106 \\ 
$^{2}$ Center for Quantum Computation and Coherence in 
Nanostructures, 
Quantum Institute, University of California, Santa Barbara, CA 93106 \\ 
$^{3}$ Department of Electrical and Computer Engineering, University 
of California, Santa Barbara, CA 93106} 
\maketitle 
\begin{abstract} 
A quantum computer is proposed in which information is stored in 
the two 
lowest electronic states of doped quantum dots (QDs). Many QDs are 
located 
in a 
microcavity. A pair of gates controls the energy levels in each QD. 
A Controlled Not 
(CNOT) operation involving any pair of QDs can be effected by a 
sequence of gate-voltage pulses which tune the QD energy levels into resonance 
with 
frequencies of the 
cavity or a laser. The duration of a CNOT operation is estimated to 
be 
much shorter than 
the time for an electron to decohere by emitting an acoustic phonon. 
\end{abstract} 

\begin{multicols}{2} 
\section {Introduction} 

A quantum computer processes quantum information which is stored in 
``quantum bits" (qubits).\cite{St98} If a small set of fundamental 
operations, or 
``universal quantum logic 
gates," can be performed on the qubits, then a quantum computer can 
be 
programmed to 
solve an arbitrary problem \cite{D85}. The explosion of interest in 
quantum 
computation can be 
traced to Shor's demonstration in 1994 that a quantum computer could 
efficiently 
factorize large integers\cite{Shor94}. Further boosts came in 1996, 
with the 
proof that quantum 
error correcting codes exist\cite{Shor95,St96}. It has since been 
shown that if the 
quantum error rate 
is below an accuracy threshold, quantum information can be stored 
indefinitely\cite{KLZ98}. 

The implementation of a large-scale quantum computer is recognized to 
be a technological challenge of unprecendented proportions. The 
qubits must be 
well-isolated 
from the decohering influence of the environment, but must also be 
manipulated 
individually to initialize the computer, perform quantum logic 
operations, 
and measure 
the result of the computation\cite{Di95}. 

Implementations of universal quantum logic gates and quantum 
computers have 
been 
proposed using atomic beams\cite{SW95}, trapped atoms\cite{PGCZ95} 
and ions\cite{CZ95}, bulk 
nuclear 
magnetic resonance\cite{GC97}, nanostructured 
semiconductors\cite{LD98,K98,BLD99,T99} and 
Josephson junctions\cite{A98,SSH97}. In schemes based on trapped 
atoms and ions, qubits couple with collective excitations or cavity 
photons. Such long-range coupling enables two-bit 
gates involving 
an arbitrary pair of qubits, which makes programming straightforward. 
However, in the 
atomic and ionic schemes\cite{PGCZ95,CZ95}, the gates must be 
performed serially, 
whereas existing 
error correcting schemes require some degree of parallelism. In 
semiconductor and 
superconductor schemes which have been proposed 
\cite{LD98,K98,BLD99,T99,A98,SSH97}, only 
nearest-neighbor qubits 
can be coupled, and significant overhead is involved in coupling 
distant 
qubits. 
However, some of these schemes have the important advantage that gate 
operations can 
be performed in parallel.

It is widely agreed that a solid-state quantum computer, if it can be 
realized, will be 
the only way to produce a quantum computer containing, for example, 
$10^3$ 
qubits. The 
remainder of this paper describes what is, to our knowledge, the 
first 
proposal for a 
semiconductor-based quantum computer in which quantum gates can be 
effected 
between 
an arbitrary pair of qubits. The qubits consist of the lowest 
electronic 
states of specially-engineered quantum dots (QDs) and are coupled by 
Terahertz cavity photons. 
The 
proposal combines ideas from the atomic and ionic implementations 
described 
above 
with recent developments in the spectroscopy of doped semiconductor 
nanostructures at 
Terahertz frequencies \cite{H95,H96,H94}. 

\section {Quantum bits and fundamental quantum logic operations}

The fundamental building blocks of the proposed computer are the 
nanostructures shown in Fig. 1. Three disks of a semiconductor 
(e.~g., GaAs) are 
embedded in a 
semiconductor with a larger band gap (e.~g., AlGaAs). The central 
disk is 
taller than the 
outer two. The barriers between the disks are sufficiently thin to 
allow 
an electron to 
rapidly tunnel between them. A structure consisting of a set of 
three 
disks and the two 
intervening barriers is hereafter called a quantum dot (QD). Each QD 
which 
is to 
participate in the quantum computation must have one and only one 
electron. 
The 
potential and four lowest electronic energy levels for a particular 
realization of a QD are 
shown in Fig. 2. The lowest two energy levels, denoted $\vert 
0\rangle$ and $\vert 1\rangle$, will 
form the qubits 
which store quantum information. The third energy level, labeled 
$\vert 2\rangle$, will 
serve as an 
auxiliary state to perform conditional rotations of the state vector 
of the 
qubit, much like 
the auxiliary state in the ion trap computer \cite{CZ95}. Below and 
above each 
QD is an 
electrical gate. Voltages applied to these gates are used to control 
the 
spacing between 
and absolute values of the energy levels of the QDs via the Stark 
effect. 
A large number 
of individually-gated QDs is contained in a 3-D microcavity whose 
fundamental 
resonance has a wavelength $\lambda_c$ much longer than a QD. A 
continuous-wave 
laser with a 
fixed wavelength different than $\lambda_c$ is introduced through one 
side of the cavity.

Fig.~3a shows the energies $E_{10}$ and $E_{20}$ of the 0-1 and 0-2 
transitions in a 
QD as a 
function of the electric field $e$ applied via the gates. Also shown 
in Fig. 
3a are the 
energies of a cavity mode photon $\ho_c$ a laser photon $\ho_l$ and 
the sum $\ho_l + \ho_c$ . The 
state of an electron in a QD can be coherently manipulated by tuning 
$E_{10}$ 
and $E_{20}$ into and 
out of resonance with $\ho_c$, $\ho_l$, and the sum $\ho_l + \ho_c$. 

A general Hamiltonian describing a QD interacting with cavity photons 
and 
the laser 
field is given by 
\begin{eqnarray} 
{\hat H} &=& \ho_c\ahat_c^+\ahat_c + E_{10}(e)\shat_{11} + 
E_{20}(e)\shat_{22}\nonumber\\ 
&+& \hbar g_{01}(e)\{ \ahat_c^+\shat_{01} + \shat_{10}\ahat_c \} + 
\hbar\Omega_{l,01}(e) \{ \shat_{01} \exp(i\omega_lt)\nonumber\\ 
&+& \shat_{10}\exp(-i\omega_lt) \} + \hbar g_{12}(e) \{ 
\ahat_c^+\shat_{12} + \shat_{12}\ahat_c \}\nonumber\\ 
&+& \hbar\Omega_{l,12}(e) \{ \ahat_c^+ \shat_{12} \exp (i\omega_lt) + 
\shat_{21} \ahat_c \exp (-i\omega_lt) \} 
\end{eqnarray} 
Here $\ahat_c$ denotes the cavity-mode annilihation operator, and 
$\shat_{ij} = \vert i\rangle\langle j\vert$ is the projection 
operator from QD state $\vert j\rangle$ to state $\vert i\rangle$. 
The vacuum Rabi frequencies are 
$g_{ij} = qz_{ij}e_{vac}$ where 
\be 
e_{vac} = \sqrt {{\ho_c\over 2\epsilon_0\epsilon V}} 
\ee 
is the amplitude of the vacuum electric field in the cavity, and 
$\epsilon$ and $V$ are the dielectric 
constant and volume of the cavity, respectively, $q$ is the 
electronic charge 
and $z_{ij}$ is the 
dipole matrix element of the $\vert i\rangle\rightarrow\vert 
j\rangle$ transition. One step in the CNOT operation 
will be 
a Rabi oscillation between states $\vert 0\rangle$ and $\vert 
2\rangle$ involving both cavity and 
laser photons at $e=e_{l+c}$. An effective Hamiltonian describing 
these two-photon processes is 
given by 
replacing the last four terms of (1) with 
\be 
H_{2-photon} = \hbar\Omega(e)\{\ahat_c^+\shat_{02} \exp(i\omega_lt) + 
\shat_{20}\ahat_c\exp (-i\omega_lt)\} 
\ee 
where the two-photon effective Rabi frequency $\tilde\Omega$ is 
given by 
\be 
{\tilde\Omega}(e) = {g_{01}(e)\Omega_{l,12}(e)\over \omega_{21}(e) - 
\omega_l} + {g_{12}(e)\Omega_{l,01}(e)\over \omega_{21}(e) - 
\omega_c} 
\ee 
The effective two-photon Hamiltonian neglects ac Stark shifts and 
terms 
which do not 
satisfy resonance conditions. In addition, we envision a scenario in 
which 
the first term 
of Eq. (4) dominates $\omega_{21}(e_{l+c}) - \omega_l << 
\omega_{21}(e_{l+c}) - \omega_c$, and the conditional phase shift 
dominates over phase shifts induced by the cavity field alone 
$(\Omega_{l,12}(e_{l+c}) >> g_{01}(e_{l+c}))$.

During the operation of this quantum computer, a qubit which is 
simply 
storing 
quantum information is in state $\vert 0\rangle$ or state $\vert 
1\rangle$, and the electric field 
across it is held at a 
fiducial value at which the energy levels of the qubit are not 
resonant 
with $\ho_c$, $\ho_l$ or $\ho_l + \ho_c$. For simplicity, we 
choose this fiducial field to be zero. For $e = e_c$, 
the first 
interaction term dominates as $\vert\omega_{10}(e_c) - \omega_c\vert 
<< \vert\omega_{10}(e_c) - \omega_1\vert$. If the cavity contains 
one 
photon or the qubit state vector is in state $\vert 1\rangle$, then 
the qubit will 
execute vacuum Rabi 
oscillations with frequency $g_{01}$, in which the probability of 
finding the 
electron in the 
excited state oscillates 90$^{\circ}$ out of phase with the 
probability of finding 
one photon in the 
cavity. For $e = e_l$, the second interaction term has the resonant 
contribution. Here, the 
state vector of the qubit rotates between states $\vert 0\rangle$ and 
$\vert 1\rangle$ with laser 
Rabi frequency $\Omega_{l,01}$. Finally, for $e = e_{l+c}$, the 
$H_{2-photon}$ dominates. If the cavity contains one 
photon and 
the qubit state vector begins in state $\vert 0\rangle$ then it 
rotates between states 
$\vert 0\rangle$ and the 
auxiliary $\vert 2\rangle$ with frequency ${\tilde\Omega}(e_{l+c})$ . 
If either the qubit is in state $\vert 1\rangle$ or the 
cavity does not contain a photon, then the qubit state vector is not 
rotated for $e = e_{l+c}$.

A controlled not (CNOT) operation is effected by a series of voltage 
pulses 
applied 
across the gates of a pair of qubits. The pulses always begin and 
end with 
the qubit at the 
fiducial electric field $(e=0)$, and rise to a target value $e_c$, 
$e_l$, or $e_{l+c}$. Figure 3b shows a 
sequence of voltage pulses which effects a two-qubit gate which is 
equivalent to a CNOT 
operation \cite{CZ95}. The cavity always begins with no photons. 
First, a 
``$\pi$"-pulse with height 
$e_c$ and duration $\pi/(2g_{01})$ is applied to the control bit. If 
the control bit 
is in state $\vert 0\rangle$, it is 
unaffected. If it is in state $\vert 1\rangle$, it rotates into 
state $\vert 0\rangle$ and acquires 
a phase $i$, and the 
cavity acquires a single photon. Next, a ``$2\pi$"-pulse with height 
$e_{l+c}$ and 
duration $\pi/{\tilde\Omega} (e_{l+c})$ is applied to the target bit. 
If the target bit is in state $\vert 1\rangle$, it is 
unaffected. If it 
is in state $\vert 0\rangle$ {\it and} the cavity contains one 
photon, it acquires a phase -1. Finally, a pulse 
with height $e_c$, identical to the first pulse, is again applied to 
the 
control bit. If there is a 
photon in the cavity it is absorbed by the control bit, returning it 
to 
state $\vert 1\rangle$ while the 
control bit acquires another phase $i$. The end result is a gate in 
which 
the state vector of 
the two-qubit system acquires a phase -1 if and only if both control 
and 
target bits are 
initially 1. The sequence of state-vector rotations which is 
effected by 
the series of 
electric field pulses is identical to the sequence effected by a 
series of 
laser pulses applied 
to cold trapped ions in Ref. [10] . In order to effect a CNOT 
operation 
(inversion of the 
target bit if and only if the control bit is 1), it is necessary to 
apply 
to the target bit ``$\pi/2$" 
and ``$3 \pi/2$" pulses with height $e_L$ and durations 
$\pi/(4\Omega_{L,01})$ and $3\pi/(4\Omega_{L,01})$, respectively, 
before and after the sequence shown in Fig. 3b [10]. 

A few additional conditions are required to ensure the fidelity of 
CNOT 
operations. 
To ensure that nearly all of the state-vector rotation occurs while 
the 
electric field is at its 
target value, the rise and fall times $\delta t$ of the pulses must 
be short 
compared to the period 
of the Rabi oscillation at the target $e$. At the same time, in 
order to 
minimize the 
probability of a transition between $\vert 0\rangle$ and 
$\vert1\rangle$ induced by the ramping 
electric field, one 
requires the changes to the Hamiltonian to be adiabatic $(\delta t >> 
\hbar/E_{10})$. As with other 
schemes for quantum computation, the timing between the successive 
pulses 
in the 
CNOT operation must be carefully adjusted to compensate for the 
quantum-mechanical 
phases accumulated by inactive qubits in their excited states. In 
order to achieve the sort of fidelity which is required for a CNOT 
operation in a 
quantum 
computer, it may be necessary to adjust the heights and durations of 
the 
electric field 
pulses to account for ac Stark shifts in the energy levels of the QDs 
which 
are induced by 
the laser field. These important details will be addressed in a 
future 
publication.

\section {Requirements for quantum computation}

The ability to effect a CNOT operation is one of several requirements 
for a universal 
quantum computer. Other requirements include:

1) Initializing the computer: Before a quantum computation begins, 
each 
qubit must 
be in a well-defined state. In the proposed computer, it suffices to 
wait, 
with all gate 
voltages at the fiducial voltage ($e= 0$) and at a temperature $T << 
E_{10}/k_B$, 
until each qubit relaxes to its ground state. For $E_{10}\approx 10$ 
meV, one requires a temperature 
$T << 120$ K. 
From calculations detailed below of the predicted energy relaxation 
times 
in QDs, a wait 
of less than 1 s will certainly ensure that all qubits are in state 
$\vert 0\rangle$.

2) Inputting initial data: At the beginning of a quantum 
computation, 
arbitrary 
rotations of the state vectors of qubits are required to load data 
into the 
qubit registers. 
Arbitrary one-bit rotations are effected using Rabi oscillations 
induced by 
the laser field, 
by applying pulses with height $e_l$ and duration between 0 and 
$2\pi/(\Omega_{l,01})$.

3) Readout: At the end of a quantum computation, the state of each 
qubit 
must be 
measured. During the read-out phase, we propose that a narrow-band 
detector with high 
quantum efficiency and the sensitivity to detect single THz photons 
be 
tuned to the 
frequency of the cavity mode $\omega_c$. The qubits can then be read 
out 
sequentially by tuning 
them to $\omega_c$. If the qubit is in state $\vert 1\rangle$, it 
will emit a photon which will 
be detected. For 
the parameters discussed below, the rate at which qubits can be read 
out 
will be roughly 
$g_{01} \geq 3\times 10^8$ Hz. A detector is required which has a 
noise equivalent power $ NEP < E_{10}/g_{01}^{1/2} \approx 10^{-17}$ 
W/Hz$^{1/2}$ with a bandwidth $> g_{01}$. While such a detector 
is not 
currently available, one of us has proposed a tunable antenna-coupled 
intersubband 
terahertz (TACIT)\cite{Ca98,Cb98} detector which will be fabricated 
from 
semiconductor 
quantum wells, could be monolithically-integrated into the quantum 
computer, and has 
been modeled to achieve the required speed and sensitivity. 

4) Error correction: Existing schemes for error correction require 
the 
execution of 
quantum logic gates in parallel. One can imagine parallelizing the 
scheme 
proposed here 
by enlarging the cavity to create several cavity modes in the 
frequency 
range over which 
QD energy level spacings are tunable. This would come at the cost of 
slowing down gate 
operations by reducing the vacuum electric field and hence the vacuum 
Rabi 
frequency $g_{01}\alpha V^{-1/2}$. A more intriguing possibility is 
to somehow marry a 
nearest-neighbor- 
coupled semiconductor scheme for quantum computation 
like\cite{LD98,K98,BLD99} with a 
nonlocal 
scheme like the one proposed here. In this case, logic gates would 
be 
effected in parallel 
in clusters of qubits coupled with nearest-neighbor interactions, 
while 
qubits in distant 
clusters could communicate serially via long-range interactions 
mediated by 
cavity 
photons.

5) Decoherence: This is the most problematic issue pertaining to 
most 
quantum 
computers. In the computer proposed here, decoherence of the 
electronic 
state of the QD 
as well as of the cavity photons must be considered. 

There are no experimental data on the decoherence of electronic 
intraband 
excitations 
in isolated QDs loaded with a single electron. Dephasing in ``open" 
quantum 
dots defined 
by gate electrodes in a 2-DEG has been studied, yielding dephasing 
times $t_{\phi} \leq 2$ ns \cite{HSMCG98}. 
The studied dots have $E_{10} \leq 20 \mu eV$. The times are 
consistent with those 
predicted for 
disordered 2-D systems. The rate of spontaneous emission of acoustic 
phonons in $\approx 200$ 
nm double QD devices containing 15-25 electrons has recently been 
deduced. 
From the 
transport currents of order $I = 10^{-12}$ A, one deduces an energy 
relaxation time of order 
$q/I = 10^{-7}$ s for transitions with energies near $50~\mu eV$ 
\cite{F98}. However, the 
transition 
energies, QD geometry, and number of electrons in the 
experiments\cite{HSMCG98,F98} are very 
different from those of Fig.~1, and it is thus impossible to draw 
conclusions about 
decoherence in the QDs envisioned in this proposal.

Many interactions will potentially cause decoherence of electrons in 
the 
computer 
proposed here. Some can be mitigated by clever engineering, 
including 

(a) the emission of freely-propagating photons which is eliminated 
because 
the QDs 
are in a 3-D cavity with a very high quality factor; 

(b) the interaction with fluctuations in the potentials of the two 
gate 
electrodes 
associated with each QD. Both cross-talk from switching voltages on 
distant QDs and 
thermal fluctuations (i.e., Johnson noise) on a QD's gate electrodes 
can 
contribute to 
fluctuating gate potentials with frequencies much lower than 
$E_{10}/\hbar$ \cite{Because,K28}. Such low 
frequency noise causes adiabatic changes $\delta E_n(t)$ to the 
energy of levels $E_n$, which lead 
phase errors $\delta\phi_n(t) = -1/\hbar \int \delta 
E_n(e_N(t^{\prime})dt^{\prime})$ (here, $n=0,1$). Such phase errors 
can be 
restricted to occur only during the time of a logic operation. The 
gate 
electrodes are made 
out of a superconductor. When a QD is not involved in a logic 
opeation, 
its two gates are connected to a superconducting ground by a 
superconducting path. Since 
there is no 
dissipation, there are no thermal fluctuations. Furthermore, 
electric 
fields generated far 
from the QD are screened by the gate electrodes. While a QD is being 
switched, the 
connection to the superconducting ground must be broken. 
Low-frequency 
noise which 
occurs during this time will contribute to an error in the accuracy 
of a 
CNOT operation. 
These and other possible errors in CNOT operation, as well as 
possible ways 
to correct 
them, will be analyzed in a future publication.

(c) the interaction with metastable traps in the semiconductor. 
Metastable 
traps in the 
semiconductor are a source of extremely slow time-varying electric 
fields 
($f<100$ Hz). If 
the fluctuating traps are far from a given gate electrode, they will 
cause 
a slowly fluctuating 
electric field near that electrode, which is screened as described in (b). 
Hence, it is only the 
traps which are fluctuating in the tiny volume between the two gate 
electrodes that pose a 
serious problem. The density of such traps in semiconductor 
nanostructures is constantly 
being reduced with advances in processing. 

(d) inhomogeneity of quantum dots. Different dots will vary 
slightly in their energy 
levels and matrix elements. This inhomogeneity can arise from 
geometrical variations 
between the quantum dots, and also from the presence of quenched 
disorder (static 
charged defects). Inhomogeneity and static disorder do not 
contribute to 
decoherence of quantum bits. However, to perform accurate one- and two-bit 
operations in an 
inhomogeneous population of quantum dots, each quantum dot in the 
quantum computer 
will need to be calibrated (by performing a CNOT operation, for 
example) before a 
quantum computation is run. Note that all solid-state 
implementations of quantum 
computers will share the need to calibrate in order to overcome 
disorder.

The lifetime of a cavity photon must be sufficiently long to enable 
many CNOT 
operations with high fidelity. This will require the development of 
few-mode THz 
cavities with extremely low loss. The expected cavity losses cannot 
be 
analyzed without 
details of the quantum computer's architecture which are beyond the 
scope 
of this paper, 
and materials properties which have not yet been measured. It is 
likely 
that cavities made 
from conventional metals will introduce losses which are 
unacceptable. One 
attractive 
possibility is dielectric cavities, made, for example from ultrapure 
Si. 
Existing 
measurements of optical loss in Si at THz frequencies are dominated 
by free 
carrier 
absorption \cite{P91}, which can be eliminated by purifying and 
cooling the Si. 
Residual losses 
in Si are due to a process in which a THz photon creates 2 phonons 
\cite{LB55}. These 
minuscule THz losses have not been measured or realistically computed 
to our knowledge. A second promising possibility is to use quantum 
dots with $E_{01}$ 
smaller than 
the energy gap of a s-wave superconductor (3 meV for Niobium, 9 meV 
for $Rb_3C_{60}$\cite{DBFZW94} 
). A cavity with volume $< < \lambda^3$ and extremely low loss could 
then be made of 
a segment of superconducting transmission line. 

\section {Times to decohere, perform CNOT}

Consider now a specific GaAs/AlGaAs QD and lossless dielectric cavity 
designed to 
minimize the time required for a CNOT operation, while at the same 
time 
avoiding the 
emission of longitudinal optical (LO) phonons ($\ho_{LO} \approx$ 36 
meV in GaAs) and also minimizing the rate of acoustic phonon 
emission. Cavity and laser photon 
energies are 
chosen to be 11.5 and 15 meV. These energies are sufficiently large 
to 
enable an 
adequate vacuum electric field $e_{vac}$ while their sum is still 
comfortably 
smaller than 
$\ho_{LO}$. Assuming perfect cylindrical symmetry, the states are 
labeled with 
quantum 
numbers $\vert l,m,n\rangle$, associated with the radial, azimuthal 
and axial degrees of 
freedom, 
respectively. The potential along the cylindrical axis of the QD 
(z-direction) and the 
numerically-computed four lowest energy levels are depicted in Fig. 
2.

Figure 3a shows the transitions $E_{10}$ and $E_{20}$ vs. electric 
field $e$. Assuming infinite walls in 
the radial direction, 
the radial wavefunctions are given by Bessel functions. The 
difference 
between the 
energy of the ground and first radial excited states is $\Delta E_r = 
30$ meV for radius $a= 13$ nm, 
assuming $m^*= m_e/15$. This is higher than the highest energy 
reached by an 
electron 
during a CNOT operation (26.5 meV = $\ho_l + \ho_c$), eliminating 
decoherence arising from coupling between axial and radial excited 
states of the QD. The growth of 
QDs similar to 
those in Fig. 1 is currently being attempted. One method is to grow 
stacked self- 
assembled QDs \cite{WLMQS97}. A second method is to make QDs made by 
growing 
GaAs/AlGaAs 
quantum wells with the conduction band profile tailored to give the 
desired 
potential in 
the z-direction (for example, that shown in Fig. 2), depositing small 
islands on top of the 
quantum well to serve as an etch mask, etching through the quantum 
well 
layers which 
are not protected by the islands, and then regrowing AlGaAs 
\cite{L98}. 

Decoherence due to the emission of longitudinal acoustic (LA) phonons 
would 
be 
difficult to mitigate given a particular QD structure. It can be 
computed 
using the 
deformation potential approximation, in which electrons scatter from 
potential 
fluctuations arising from local volume compressions and dilations 
induced 
by LA 
phonons. Piezoelectric coupling between electrons and transverse 
acoustic 
phonons 
exists in III-V semiconductors, but is thought to be weaker than 
deformation potential 
coupling \cite{SNCH96,OT89}. 

Following Bockelmann \cite{B94}, assuming zero temperature, the rate 
at which 
electrons 
relax between QD states by emitting LA phonons is given by Fermi's 
golden 
rule 
\be 
\tau^{-1}_{i\rightarrow f} = {2\pi\over\hbar} \, \sum_i\, 
\left|\langle\psi_1\vert W\vert\psi_f\rangle\right|^2\, \delta(E_f - 
E_i - E_{\bar k}) 
\ee 
Here, the deformation potential interaction is given by 
\be 
W = \sqrt {{\hbar K\over 2\rho c_i}} \, De^{i{\bar k}\cdot {\bar 
x}}\, , 
\ee 
where $K = \vert {\bar k}\vert$, $\rho = 5300$ kg/m$^3$, $c_s= 3700$ 
m/s, and the deformation potential $D= 8.6$ eV. 
We 
approximate the eigenfunctions associated with motion in the 
$z$-direction by 
those for an 
infinite square well of the width 40 nm, which fit the exact 
wavefunctions 
reasonably 
well. As the volume $V$ of the crystal is taken to infinity, it can 
be shown 
that the 
expression for $\tau_{10}$ becomes 
\begin{eqnarray} 
&\tau_{10}^{-1} = {D^2K^3_{10}\over 4\pi\hbar\rho c^2_s}\, \int^1_0\, 
dq^{\pr}\, {q^{\pr}\over \sqrt{1 - q^{\pr 2}}}\nonumber\\ 
&\left| N\int^{x_{01}}_0\, dr^{\prime} J_0(\alpha q^{\pr} 
r^{\pr})J_0^2 (r^{\pr})r^{\pr}\right|^2\nonumber\\ 
&\times\left| {2\over \pi}\, \int^{\pi/2}_{-\pi/2}\, 
dz^{\pr}\cos(z^{\pr})\sin(2z^{\pr})e^{i\beta\sqrt{1-q^{\pr 2}} 
z^{\pr}}\right|^2 
\end{eqnarray} 
Here, $K_{10} = E_{10}/\hbar c_s$, $q^{\pr} = q/K_{10}$, where $q$ is 
the radial phonon wave-vector, $N^{-1} = \int^{x_{01}}_0\, dr^{\pr} 
r^{\pr} J^2_0(r^{\pr}) = 0.779$ is the normalization factor for the 
radial wavefunctions, $x_{01} = 2.404$ is the first zero of $J_0(a)$, 
$r^{\pr}= rx_{01}/a$, where $r$ is the radial distance, $z^{\pr}= 
z\pi/h$, $h=40$ nm is the height of the QD, $\alpha = 
K_{10}a/x_{01}$, and $\beta = K_{10}h/\pi$. At $e=0$, one finds 
$E_{10}=12.25$ meV,$K_{10}= 5.03$ nm$^{-1}$, $\alpha = 27$, and 
$\beta = 64$. The relatively large values 
of $\alpha$ and $\beta$ 
indicate that the characteristic phonon wavelength is much shorter 
than the 
size of the 
QD, leading to very small values for the integral over $q^{\pr}$ 
\cite{ZR98}. The value 
of $\tau$ computed by 
evaluating Eq. (7) numerically at the above parameters is $150~\mu$s.

The time required to execute a CNOT operation for the particular QD 
structure is now 
estimated \cite{gate}. For a dielectric cavity resonating at $\ho_c 
= 11.5$ meV with 
index of 
refraction $n= 3.6$, the maximum vacuum electric field 
$e_{vac}\approx 49$ V/m is achieved 
for a 
cavity with minimal volume $(\lambda_c/2)^3$, where $\lambda_c = 
c/n\omega_c = 30~\mu$m is the wavelength 
of the 
resonant radiation inside the cavity. This vacuum electric field, 
together 
with the matrix 
elements $z_{10}$($e_c= 1.177$ MV/m)= 6.32 nm, $z_{10}(e_{l+c}= 
0.7668$ MV/m) = 6.95 nm, $z_{10}(e_l = 1.682$ MV/m) = 4.97 nm, and 
$z_{21}(e_{l+c}= 0.7668$ MV/m)= 6.52 nm, and a laser electric field 
of 30.7 
kV/m, enable one to compute the time required for a CNOT operation. 
The $2\pi$ 
pulse 
applied to the target bit requires interaction with both a laser and 
a 
cavity photon, and 
hence is by far the longest operation, requiring 25 ns. The $\pi$ 
pulses 
applied to the control 
bit require 3.3 ns each. Unconditional 1-bit rotations which occur 
at $ e= 
e_l$ 
take only a few 
ps for a laser electric field of 30 kV/m. It is likely the laser 
would 
need to be attenuated 
for these rotations, in order to satisfy the requirement that the 
transition time for the 
electrical pulse is much shorter than the period of the Rabi 
oscillation at 
the target electric field. If the only mechanism for decoherence is 
given by acoustic phonon 
emission, then 
the above calculations suggest that several thousand CNOT operations 
can be 
performed 
before the computer decoheres.

\section {Conclusions}

We have proposed a quantum computer in which quantum information is 
stored 
in the 
lowest electronic levels of doped quantum dots. The energy levels in 
each 
dot are 
controlled by dedicated gate electrodes. THz photons in a cavity act 
as a 
data bus which 
can couple an arbitrary pair of quantum dots. A sequence of 
adiabatic 
voltage pulses 
applied to individual quantum dots can effect a CNOT operation 
involving 
any two 
quantum bits in the computer. We hope that this concrete proposal 
for a 
quantum 
information processor will stimulate theoretical and experimental 
activity.
As with all proposals for quantum computation, the obstacles to 
implementing this 
one are formidable. Among the most important challenges, new types 
of QDs 
must be 
constructed, gated and loaded with single electrons \cite{Qd}; 
few-mode THz 
cavities with 
extremely high Q must be fabricated; and single THz photons must be 
detected. 
Although each of these worthy challenges is beyond today's state of 
the 
art, the rapid pace 
of progress in materials science and THz technology makes us 
optimistic 
that these 
obstacles will be overcome in the not-too-distant future. 

\acknowledgments
We gratefully acknowledge Profs. D. D. Awschalom, John Davies, 
Matthew 
Fisher, 
Art Gossard, Evelyn Hu and Horia Metiu, and Mr. Wyatt Wasicek for 
useful 
discussions. 
This work was supported by ARO DAAG55-98-1-0366 and a David and 
Lucile Packard Fellowship (A. I.).

\narrowtext
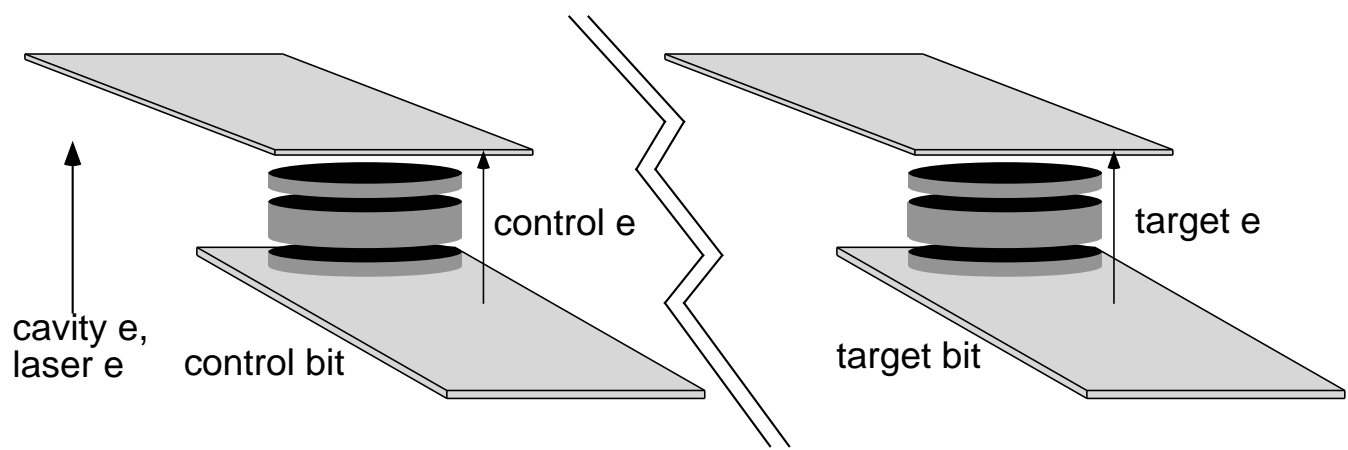
\begin{figure} 
\caption{Fundamental elements of the proposed quantum computer. Each 
set of quantum dots (QDs) contains 1 electron, and is individually 
addressable by 
a pair of gate 
electrodes. One QD is chosen to be a control bit, the other a target 
bit 
for a controlled-not 
(CNOT) operation. Many fundamental elements are embedded in a 
single-mode 
cavity.} 
\label{figureone} 
\end{figure}

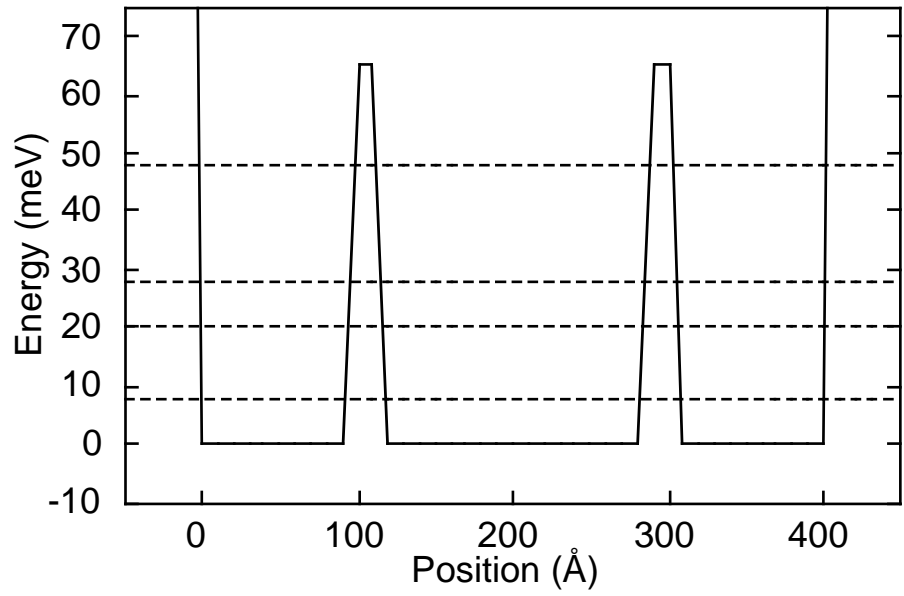
\begin{figure} 
\caption{Potential and energy level diagram for the lowest energy 
levels of 
a set 
coupled QDs which is suitable for a qubit. The ground ($\vert 
0\rangle$) and first 
excited ($\vert 1\rangle$) states 
are used to store quantum information. The second excited state 
($\vert 2\rangle$) is 
an auxiliary state 
which is used to effect a controlled not operation, but does not 
store 
quantum 
information. The height of the QD is 41 nm, and the potential inside 
is 0 
except for two 2 
nm barriers with 65 meV potential which separate the central 17 nm 
well 
from the outer 
10 nm wells. } 
\label{figuretwo} 
\end{figure}

\begin{figure} 
\caption{(a) Transition energies between states $\vert 0\rangle$ and 
$\vert 1 \rangle$ ($E_{10}$) and between $\vert 0\rangle$ and 
$\vert 2\rangle$ ($E_{20}$) vs applied electric field, and photon 
energies of a cavity mode ($\ho_c$), a laser 
($\ho_l$ ), and the sum $\ho_l + \ho_c$ . The $E_{10}$ transition 
resonates with $\ho_c$ and $\ho_l$ at electric fields $e_c$ and 
$e_l$, respectively. The $E_{20}$ transition resonates with the 
two-photon transition 
with energy at electric field $e_{l+c}$. 
(b) A sequence of electric field pulses to a control and a target 
bit 
which are used in a 
CNOT gate. First, a ``$\pi$" pulse is applied to the control bit, 
transferring 
a photon to the 
cavity and multiplying the state vector by $i$ if and only if the 
control 
bit is 1. Then, a 
``$2\pi$" pulse is applied to the target bit, multiplying the state 
vector by -1 
if and only if 
there is a photon in the cavity and the target bit is in its ground 
state. 
Finally, a second 
``$\pi$" pulse is applied to the control bit, removing the photon 
from the 
cavity, returning the 
control bit to the excited state, and again multiplying the state 
vector by 
$i$. The state 
vectors in which the control bit is 0 are unaffected by the sequence 
of 
electric field pulses, 
and thus are not shown. One-bit rotations can be effected by applying 
an 
appropriately- 
timed pulse with amplitude $e_l$. As shown by Cirac and Zoller [10], 
the gate 
shown here, together 1-bit rotations on the target bit, result in a 
CNOT operation.} 
\label{figurethree} 
\end{figure} 
\end{multicols} 

\begin{references}
\bibitem{Because} Because the proposed computer will be operated at 
temperatures $T<< E_{10}/k_B\approx 100$ K, and Johnson noise can be 
viewed as the low-frequency limit of black-body radiation[26], any 
Johnson noise in this computer will be at frequencies $\omega << 
E_{10}/\hbar$.
\bibitem{gate} The gate electrodes can be engineered to act as 
antennas, enhancing the 
vacuum 
electric field at the QDs and hence shortening the time required for 
a CNOT 
operation. This effect will be considered in a future publication.
\end{references}
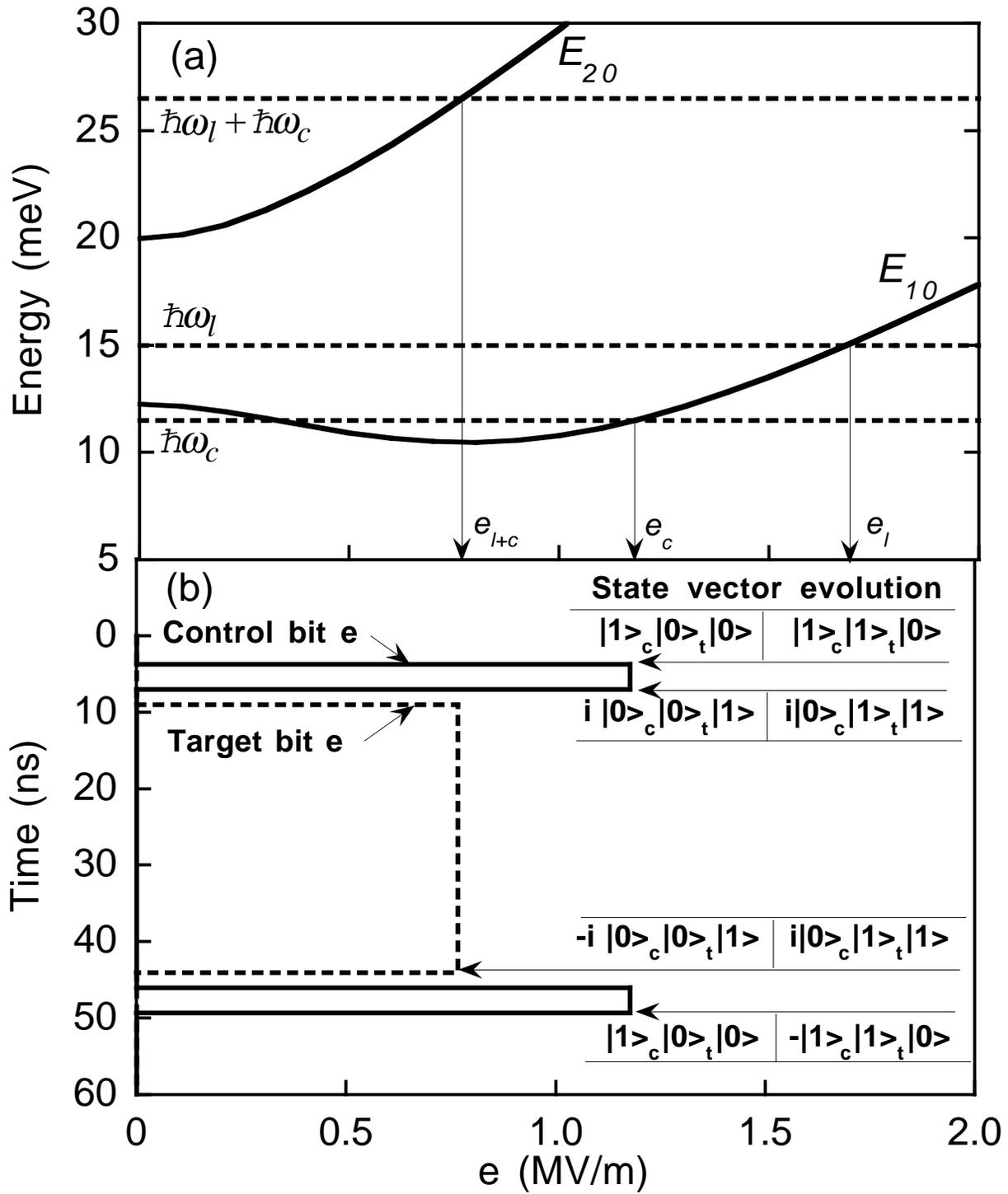
\end{document}